\documentclass[3p, 11pt,authoryear]{elsarticle}
\usepackage{setspace}
\usepackage{graphicx} 
\usepackage{amsmath}
\usepackage{circuitikz}
\usepackage{adjustbox}
\usepackage{array}
\usepackage{amssymb}
\usepackage{optidef}
\usepackage[hyphens]{url}

\usepackage{natbib}
\usepackage{lipsum}

\usepackage{subcaption}
\usepackage{caption}
\journal{Elsevier}
\begin{document}

\begin{frontmatter}

\title{The impact of heatwave-driven air conditioning adoption on electricity demand: A spatio-temporal case study for Germany}

\author[1,2]{Leo Semmelmann}
\ead{leo.semmelmann@kit.edu}
\author[2]{Frederik vom Scheidt}

\address[1]{Karlsruhe Institute of Technology, Karlsruhe, Germany}
\address[2]{Quantensprung Energy Consulting, Munich, Germany}

\date{August 2025}

\begin{abstract}
Intensifying heatwaves driven by climate change are accelerating the adoption of mobile air conditioning (AC) systems. A rapid mass adoption of such AC systems could create additional stress on electricity grids and the power system. This study presents a novel method to estimate the electricity demand from AC systems both at the system level and at high temporal and spatial granularity. We apply the method to a near-future heatwave scenario in Germany in which household AC adoption increases from the current 19\% to 35\% during a heatwave similar to the one of July 2025. We analyze the effects for 196,428 grid cells of one square kilometer across Germany, by combining weather data, census data, socio-demographic assumptions, mobility patterns, and temperature-dependent AC activation functions. We find that electricity demand of newly purchased mobile AC systems could increase the peak load by over 12.9 GW, with urban hot-spots reaching 5.2 MW per square kilometer. The temporal pattern creates a pronounced afternoon peak that coincides with lower photovoltaic generation, potentially exacerbating power system stability challenges. Our findings underscore the urgency for proactive energy system planning to manage emerging demand peaks.
\end{abstract}


\end{frontmatter}

\clearpage
\section{Introduction}

Climate change is increasing the likelihood and intensity of heatwaves. Such heatwaves in turn lead to a higher mortality risk for affected individuals \citep{fischer2010consistent, liu2024rising, heat-deaths-2025}. It has been shown in literature that households react to heatwaves by adopting air conditioning (AC) devices \citep{zhang2024climate}. Compared to fixed installations of AC systems, mobile AC systems can be quickly installed at limited costs (mostly only a few hundred euros). For example, during the recent heatwave in Europe, consumers bought new mobile AC systems in large numbers, which were soon sold out in some German cities \citep{Galaxus_2025, nn2025klimageraete, swr_2025}.

A disruptively quick adoption of such mobile AC systems, fueled by heatwaves, the aging European population, and growing urbanization could boost electricity demand with potentially severe effects on the electricity system. The uptake of AC systems due to heatwaves has been identified as a key driver for the substantial increase in global electricity demand in the building sector \citep{IEA}.  A past study estimates that the uptake of AC systems creates 34 TWh of new electricity demand per year in Europe \citep{colelli2023air}. This indicates the substantial effect of AC devices on electricity systems. For forward-planning policymakers, regulators, and distribution grid operators, it is critical to understand when and where demand peaks will occur, and therefore, an analysis on a more granular temporal and spatial level is opportune.

Therefore, this paper makes two key contributions. First, it presents a novel method for estimating the potential impact of rapidly rising AC adoption on electricity demand at a granular level using demographic data. Secondly, using this method, it quantifies the effects for a near-future scenario in Germany.

To achieve this, this paper proceeds as follows. In Section \ref{sec_related}, related literature is reviewed and a prevailing research gap is identified. Section \ref{sec_method} mathematically describes a novel, widely applicable method for AC-driven electricity demand estimation at high temporal and geographical resolution. Section \ref{sec_data} provides a case study for mobile AC adoption in Germany based on near-future adoption estimates as well as highly granular census and weather data. Section \ref{sec_results} presents the key results of this case study, i.e., the expected hourly electricity demand from mobile AC systems per 1km\textsuperscript{2} census grid cell. In Section \ref{sec_discussion} we discuss these findings, limitations and important implications for policy makers and grid operators. Section \ref{sec_conclusion} summarizes and concludes.

\section{Related work} \label{sec_related}

The related work can be structured into two main streams: climate-energy models of cooling demand and behaviorally driven AC usage models.

Climate-energy models of cooling demand include top-down statistical approaches and bottom-up building-stock modeling, often implemented as urban building energy models.
Top-down approaches link aggregated cooling demand to weather indicators such as cooling degree days and macro-level socio-economic drivers. These models are computationally efficient and widely used for national or regional forecasts, but they lack the ability to capture fine-scale spatial variation in demand.
Bottom-up and urban building energy models typically simulate cooling demand using building physics, technology performance, and archetype-specific characteristics. They thus provide high spatio-temporal details but also require very detailed inputs \citep{Ferrando_2020}.
Recent advances extend these methods to large geographies. In particular, the Demand.ninja framework and related paper by \cite{ninja_2023} provide estimates of heating and cooling demand on multiple spatial scales, validated on more than 5000 buildings around the world. The model considers temperature, solar irradiance, wind speed, and humidity to estimate indoor temperatures in buildings. The model uses cooling degree days and a linear regression fitted on empirical data to estimate the daily electricity demand from AC systems. They find that global warming increases the cooling demand in Europe by up to 5\% per year and that AC systems in the US consumed 66 TWh/a more in 2020 than in 1980 due to hotter summers. However, the study does not explicitly model demographic-specific occupancy probabilities that influence when residents are at home and able to operate AC units. This omission might limit the model's accuracy in predicting regional peak hours of residential AC demand \citep{liu2024exploring}. In their review article of urban building energy models, \cite{DAHLSTROM2022112099} identify stochastic and dynamic occupancy modeling and the integration of socio-economic data as important prevailing research gaps.

Behaviorally driven AC usage models explicitly model occupant behavior and device activation patterns.
Probabilistic activation functions predict switch-on events based on, e.g., outdoor temperature and other environmental triggers \citep{ren_2014}. Stochastic behavioral models (e.g., Markov chains) incorporate occupant schedules to refine the temporal distribution of cooling loads \citep{Gunay02012018, Japan_2022}. Such models can achieve high temporal realism, which is critical for capturing short but pronounced demand peaks. However, they are often applied at the building or neighborhood level and are rarely embedded into grid-scale AC demand models, which are relevant to power system operators. \cite{liu2024exploring} introduce an approach to integrate occupant behavior modeling into urban building energy modeling, focusing on heterogeneity and stochasticity. For a case study of 1,000 households in Hangzhou, China, they find that heterogeneity is the main reason for the observed diversity in cooling demand.

Both streams of research contribute essential modeling ingredients, but few studies integrate them at both high spatial and temporal resolution across an entire country. Climate-energy models often scale nationally, but smooth out spatial variability. Behavioral models capture timing but rarely scale beyond city or neighborhood studies.
This gap is especially critical for distribution grid and power system planning, where high-resolution demand profiles are needed to identify hotspots and manage short-term load surges during extreme heat events. Our work addresses this gap by 
a) applying probabilistic AC activation functions informed by socio-demographic occupancy schedules, 
b) mapping demand across a uniform 1 km² national grid for Germany, revealing local load concentrations, 
c) quantifying short-term peak load impacts during a heatwave-driven rapid adoption scenario. Thereby, we aim to provide results that are directly actionable for grid operators and policymakers in the context of climate-driven AC demand growth.

\section{Methodology} \label{sec_method}
We estimate the additional electricity demand from mobile air conditioning systems using a spatially and temporally disaggregated approach. Our methodology combines census data, socio-demographic assumptions, mobility patterns, and temperature-dependent AC activation functions to calculate hourly expected demand for each grid cell.\\

\textbf{Spatial framework:} First, we describe the spatial resolution of our target country as equally-sized grid cells. Let $G$ denote the set of all grid cells. For each grid cell $g \in G$, census data is used for the number of households of different sizes \citep{zensus2022_population}:
\begin{equation}
H_{g,s} = \text{number of households with } s \text{ persons } \text{ in grid cell } g
\end{equation}
where $s \in \{1, 2, 3, 4, 5, 6+\}$ represents household sizes.
Each grid cell $g$ is assigned to the nearest weather station using k-means clustering, creating a mapping function $\phi: G \rightarrow W$, where $W$ is the set of weather stations.\\

\textbf{Demographic model:} Second, we map the different observed household sizes to assumed household types. This is required to develop different AC activation profiles, depending on household occupancy. Hence, we define a socio-demographic distribution matrix $\mathbf{\Pi} \in \mathbb{R}^{6 \times 5}$ where element $\Pi_{s,d}$ represents the probability that a household with $s$ persons belongs to demographic group $d$:
\begin{equation}
\Pi_{s,d} = \Pr(\text{household type } = d \mid \text{household size } = s)
\end{equation}
where $d \in D = \{$families, couples without children, retired, shared flats, singles$\}$ and the probability constraint holds:
\begin{equation}
\sum_{d \in D} \Pi_{s,d} = 1 \quad \forall s \in \{1, 2, 3, 4, 5, 6+\}
\end{equation}
The total number of households of demographic group $d$ in grid cell $g$ is:
\begin{equation}
H_{g,d} = \sum_{s} H_{g,s} \cdot \Pi_{s,d}
\end{equation}

\textbf{Mobility and home presence modeling:} The probability that at least one person from a household of type $d$ is at home at hour $h$ is denoted as, oriented on \cite{flett2021modelling}:
\begin{equation}
\alpha_{d,h} = \Pr(\text{person at home at hour } h \mid \text{demographic group } = d)
\end{equation}
where $h \in \{0, 1, 2, \ldots, 23\}$ represents hours of the day and $\alpha_{d,h} \in [0,1]$ for all $d, h$.

We note that this is a significant simplification of household mobility behavior, as such data is rarely available. For example, it assumes that the time-dependent probability of being at home for a family with three household members equals that of a family with six or more inhabitants. In addition, the probabilities of a 20-year-old single are assumed to be the same as those of a 50-year-old single. This simplification allows the model to work with publicly available data and to easier be transferred to other geographies.\\

\textbf{Temperature-dependent AC activation function:} The AC activation probability depends on temperature $T$ through a discrete three-parameter Weibull cumulative function, formalized as, derived from \cite{liu2024exploring}:

\begin{equation}
p(T) = 
\begin{cases}
0, & \text{if } T < u \\[6pt]
1 - \exp \left\{ - \left[ \frac{T - u}{l} \right]^k \left( \frac{\Delta t}{\tau_c} \right) \right\}, & \text{if } T \geq u
\end{cases}
\end{equation}

Here, $u$ is the threshold temperature (°C), which represents the lowest temperature at which there is a probability that occupants will switch on the AC. $l$ is the scale parameter (°C) used to non-dimensionalize the temperature. $k$ is a shape parameter that describes the sensitivity to the environment. $\Delta t$ refers to the time step of the data, and $\tau_c$ is the time constant.\\

\textbf{Expected demand calculation:} For each grid cell $g$ and hour $h$, we calculate the expected hourly additional electricity demand $E_{g,h}$ from AC systems by combining all probabilistic factors in a single equation:

\begin{equation}
\mathbb{E}[E_{g,h}] = \sum_{d \in D} H_{g,d} \cdot \alpha_{d,h} \cdot p(T_{\phi(g,h)}) \cdot P^{\max} \cdot \Delta^T \cdot \eta
\end{equation}

This formulation multiplies the number of households of a demographic group $d$ in grid cell $g$ by the probability $\alpha$ that someone from that demographic group is home at hour $h$, the temperature-dependent AC activation probability $p$ at the corresponding weather station (with measured temperatures $T$), the maximum AC power consumption per unit $P^{\max}$, the time resolution of our simulation $\Delta^T$ and the overall adoption rate of AC systems $\eta$. 

Note that since $\Delta^T = 1$ hour in our analysis, the resulting energy values $E_{g,h}$ are numerically equivalent to power values, and these terms are used interchangeably throughout this work.

The expected national additional electricity demand at hour $h$ is obtained by summing over all grid cells:
\begin{equation}
\mathbb{E}[E_h^{\text{national}}] = \sum_{g \in G} \mathbb{E}[E_{g,h}]
\end{equation}

\section{Data} \label{sec_data}

Our analysis is based on various data sources. We assume an overall long-term AC adoption rate of 35\%, in line with the European average expected by the European Commission for 2030 \citep{eu_commission_ac_estimates_2030}. Since approximately 19\% of households already have AC systems today \citep{verivox2024klimaanlagen}, this corresponds to an additional adoption of 16\%-points. The impact of this increase in mobile AC adoption on the German power system is then modelled on a 1km\textsuperscript{2} grid cell resolution. The underlying data is based on recent German census data.

\subsection{Census data}

Our analysis leverages detailed demographic and housing information from the 2022 German census, which provides high-resolution population data for $|G| = 196{,}428$ grid cells at a $1\text{km}^2$ spatial level \citep{zensus2022_population}. The Zensus 2022 was conducted as a register-based census by the German Federal Statistical Office, supplemented by a representative household survey and a building and housing census. It gathered essential statistics on the population of Germany, including their living conditions and workplaces. 

\begin{figure*}[htbp]
\centering
\adjustbox{max width=\columnwidth}{
\includegraphics{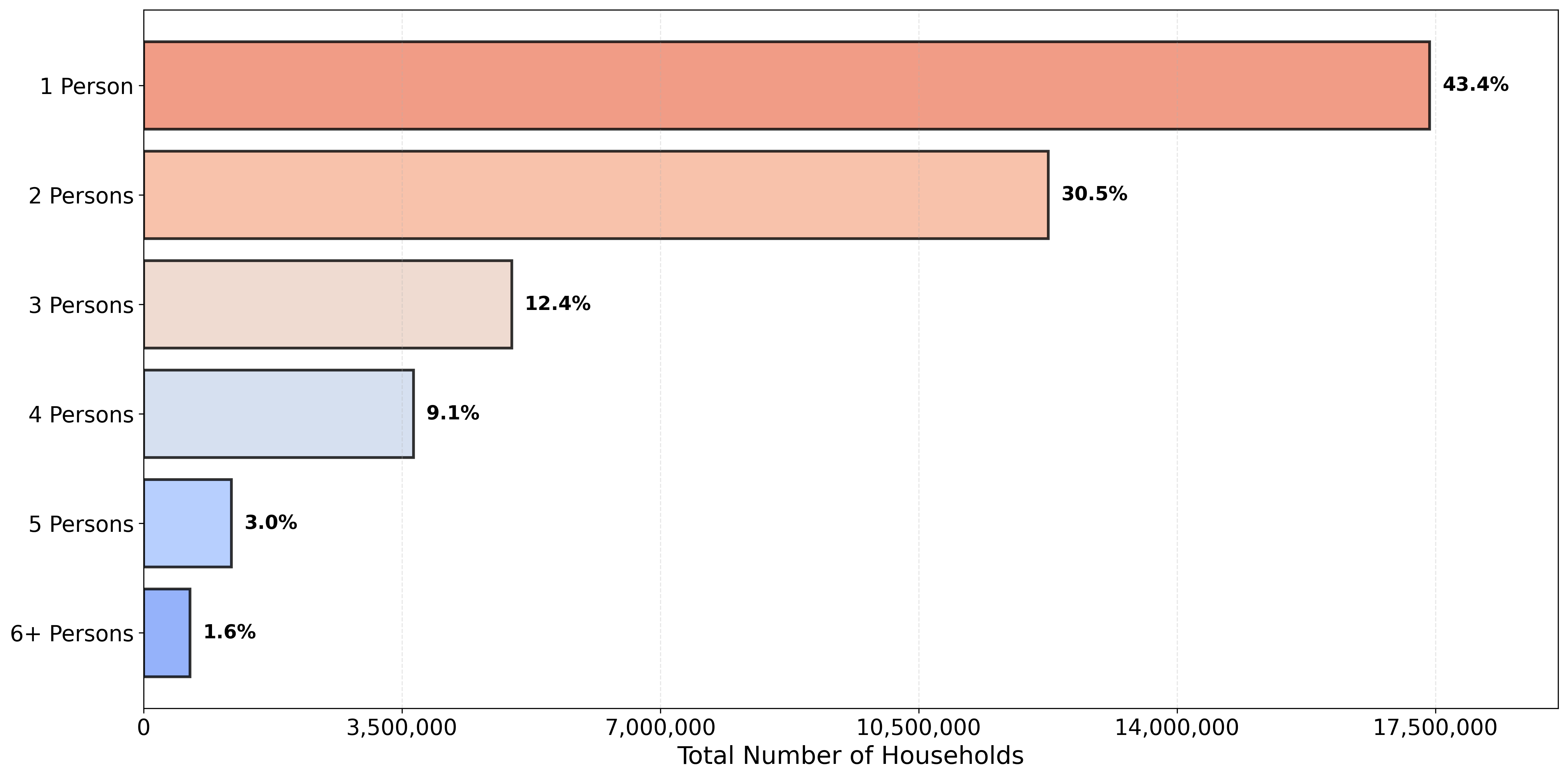}
}
\caption{Distribution of household sizes, according to census data.}
\label{fig:households_distribution}
\end{figure*}

One major information we draw from the census data is the sizes of households (from 1 Person to 6+ Persons), from which we later infer their household types (singles, families, retired, etc.). In total, we consider 40,138,344 households, with one-person households having the highest share (43.4\%), as depicted in Figure \ref{fig:households_distribution}.

In a following step, we infer the distribution of household types per household size (i.e., when there is a 1-person household, what is the probability that it is a retired individual) based on census data \citep{zensus2022_table2081} and pensioners statistics \citep{zensus2022_table1035}. We describe the resulting socio-demographic distribution model in greater detail in Table \ref{tab:household_composition} in the Appendix. We note that we estimate a uniform distribution of household types per household size across Germany, since the publicly available census data does not specify the demographic composition of household sizes. 

\subsection{Mobility data}

Subsequently, the inferred household types are mapped onto different probabilities of being present at home, which is a precondition for activating the AC. 

\begin{figure*}[htbp]
\centering
\adjustbox{max width=\columnwidth}{
\includegraphics{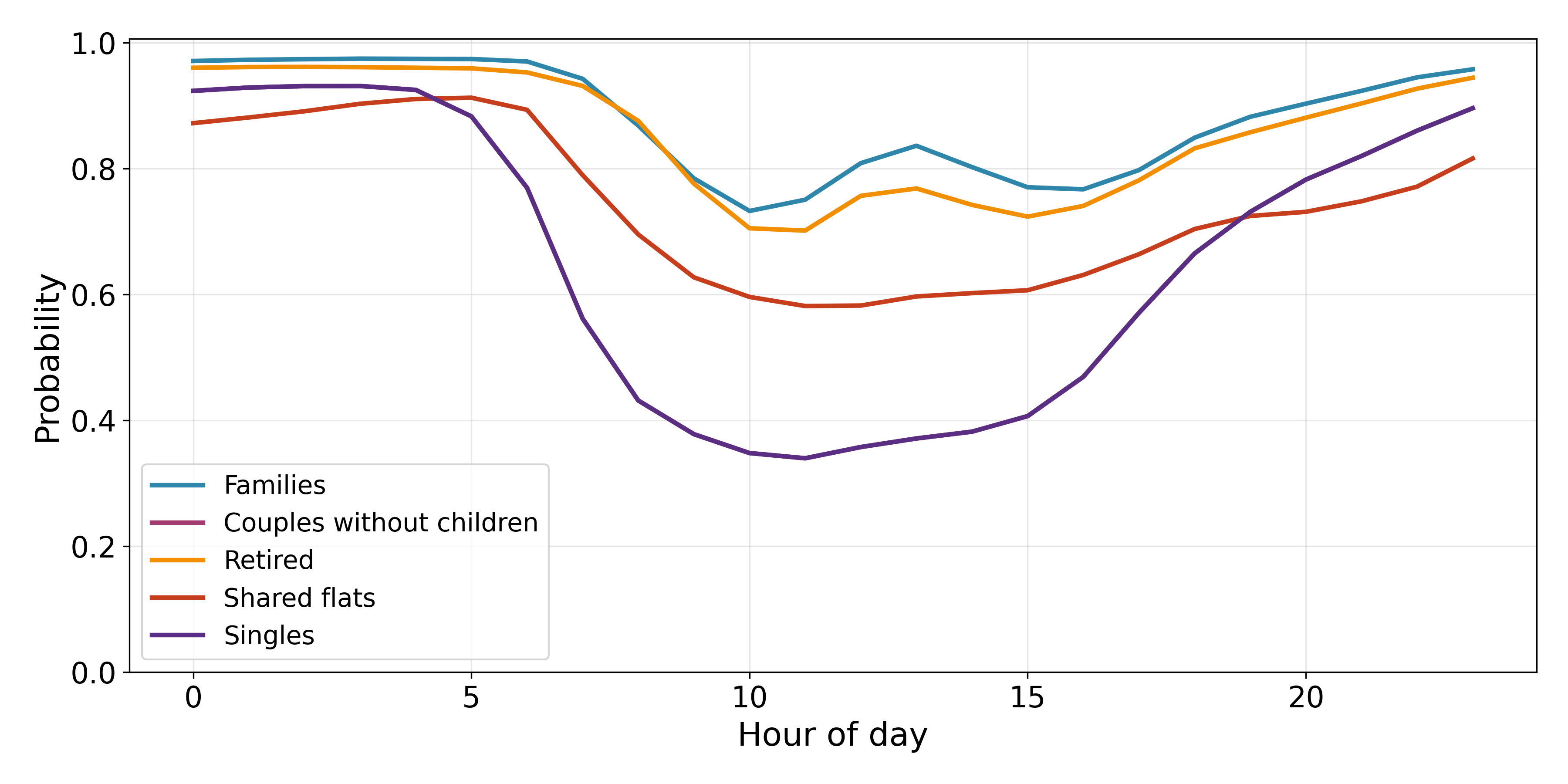}
}
\caption{Hourly home presence probabilities by demographic group throughout a typical day}
\label{fig:home_probability}
\end{figure*}

To this end, we use data from the German mobility panel, as provided by \cite{huber_2019}. That data set contains mobility data from 6.465 car users over the course of one week at 15-minute resolution. Namely, it includes information about when a person is at home and the (main) occupation of each person. For the sake of this study, we refactor the time series to hourly resolution and calculate the \textit{home presence probability} across all persons and days for each occupation group. Then, we map each census household type to a mobility type, or a combination of multiple mobility types. For Singles and Couples, we use \textit{Full-time worker} mobility profiles. For Families, we use the larger value of \textit{Full-time worker} and \textit{Homemaker} in each hour. For Retired we use the \textit{Retired} profile. For Shared Flats, we use the larger value of \textit{(Higher) Education} and \textit{Apprenticeship} in each hour. The resulting home presence probability curves are depicted in Figure~\ref{fig:home_probability}.

\subsection{AC power and activation function}

We uniformly set the AC maximum capacity at 2.1 kW, based on a frequently sold mobile AC model \citep{delonghi_pac_es72_young}, for which we assume binary operation (either off or at full power). We note that compared to cheaper models, the used technology is rather on the more efficient side of AC systems, with an efficiency label of A. The AC activation probability depends on temperature through a discrete three-parameter Weibull cumulative function from \cite{liu2024exploring}. The function is parametrized with threshold temperature $u = 18.5^\circ$C, scale parameter $l = 3.5^\circ$C and shape parameter $k = 3.5$ and exemplarily depicted in Figure \ref{fig:heterstoactiv} in the Appendix. We follow \cite{ninja_2023} in arguing that opening windows during hot weather (e.g., during the night when outdoor temperature is cooler) has a resetting effect on indoor temperature and thus outdoor temperature is an adequate descriptor for AC activation in residential buildings.

\subsection{Weather data}

For modelled households that we assume to be at home, the temperature-dependent activation function decides whether AC systems are operated or not. 

For an accurate modelling of weather data in every grid cell, we retrieve weather data from 100 weather stations distributed across Germany (i.e., $|W| = 100$) from \citep{visualcrossing_weather}. Then, based on a k-means clustering process, we map every grid cell to the closest weather station and use the connected temperature profile as input for the AC activation function. The exact locations of the considered weather stations are shown in Figure \ref{fig:weather_stations} in the Appendix. In Figure \ref{fig:weather_data}, we depict the daily temperature curves on the 2nd of July, 2025, for all considered weather stations. Observed temperatures range between 13.3°C and 37.6°C; the average daily temperature over all weather stations lies at 26.2°C, underlining the presence of a heatwave on the investigated day. We note that between weather stations, we can see heavy differences in temperatures: the median variation between the lowest and the highest temperature per hour lies at 17.5°C. This underlines the importance of geospatially modelling the power system impact of AC adoption, since temperature-dependent activation per grid cell can vary heavily.

\begin{figure*}[htbp]
\centering
\adjustbox{max width=\columnwidth}{
\includegraphics{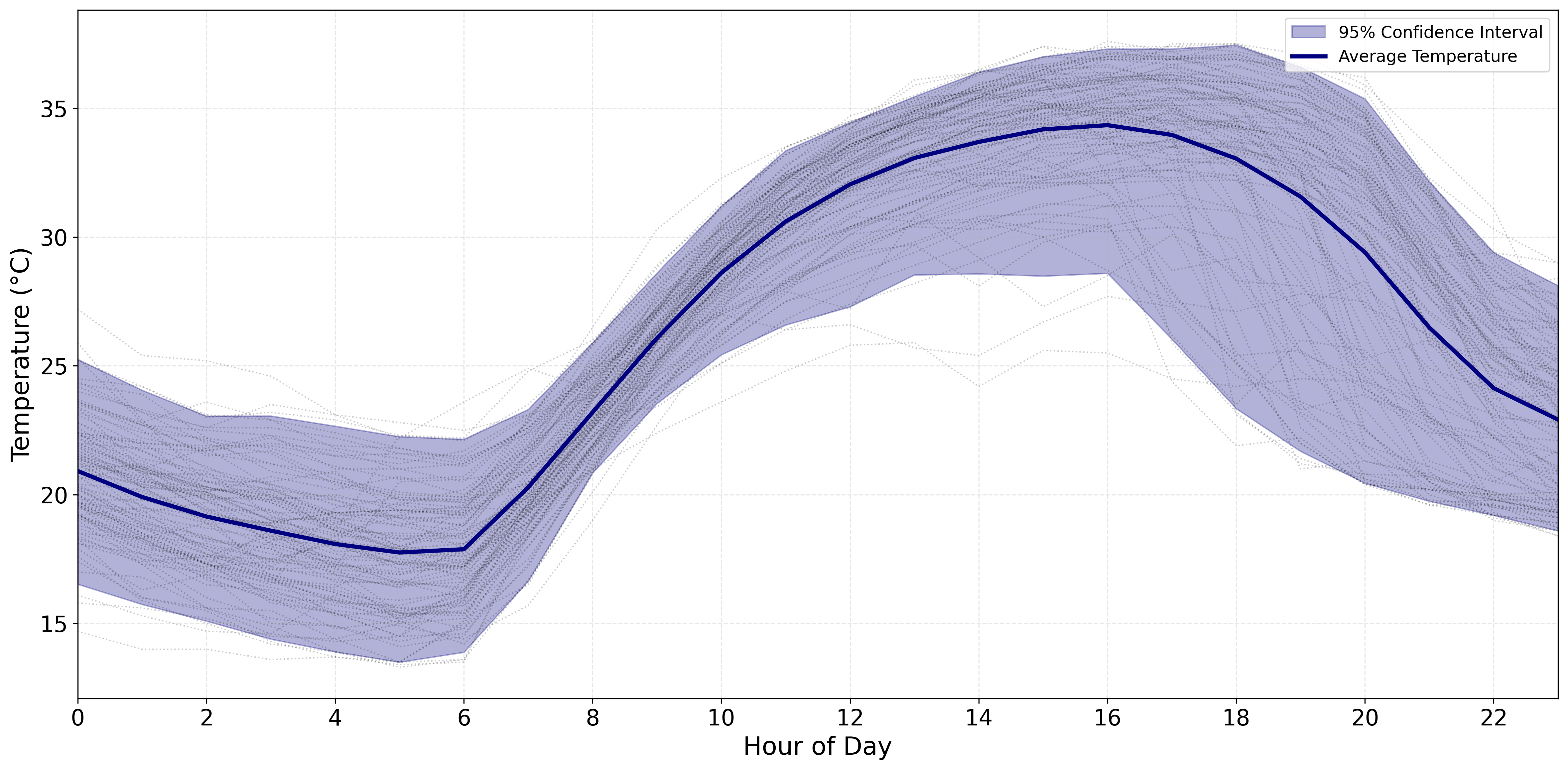}
}
\caption{Temperature data of the 100 considered German weather stations for the 2nd of July, 2025.}
\label{fig:weather_data}
\end{figure*}

\section{Results} \label{sec_results}

Our analysis of potential additional electricity demand from mobile air conditioning systems in Germany reveals several critical findings for energy system planning. 

\begin{figure}[htbp]
\centering
\adjustbox{max width=\columnwidth}{
\includegraphics{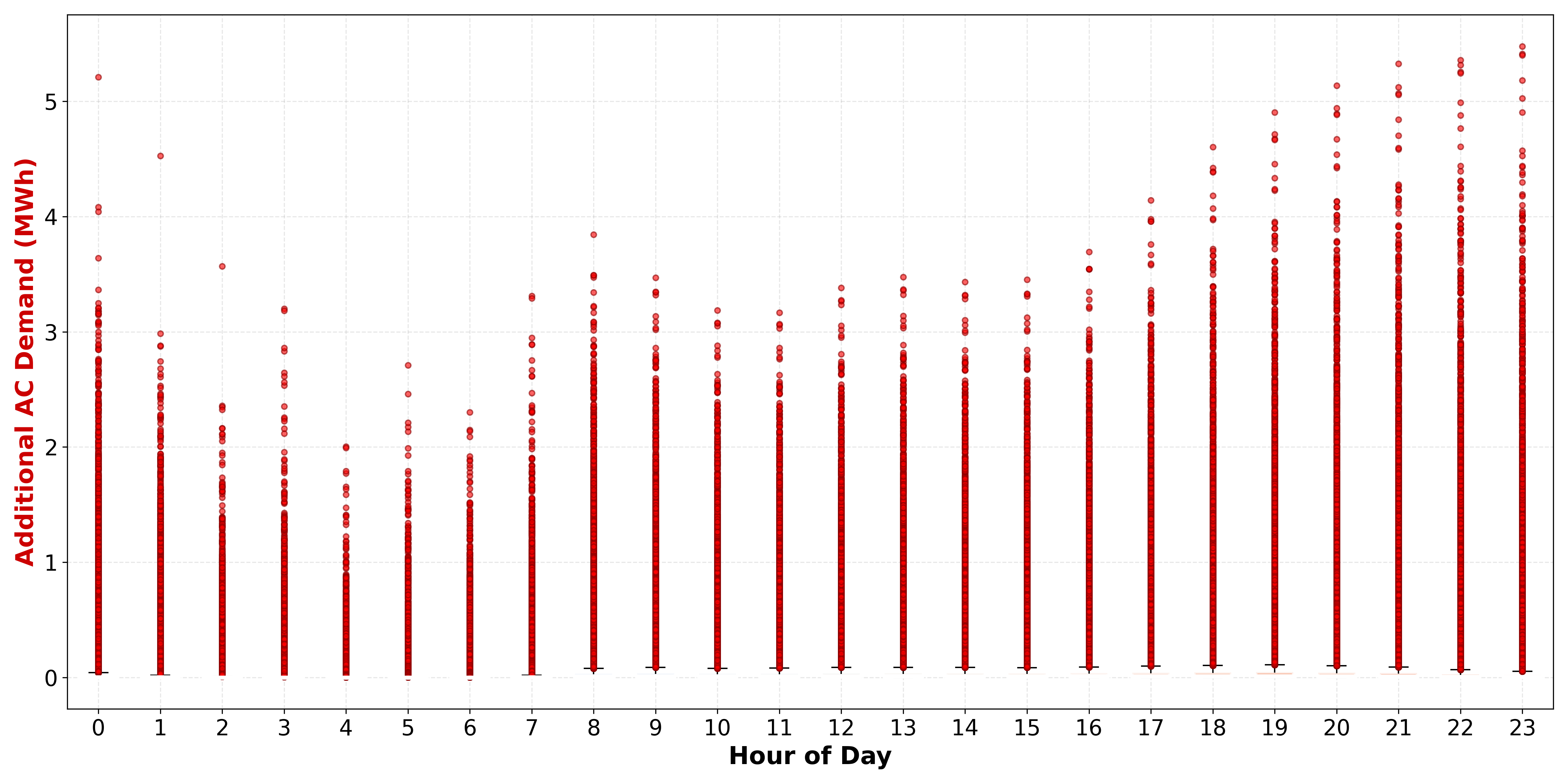}
}
\caption{Hourly distribution of expected additional loads through AC installations across all investigated 1km\textsuperscript{2} grid cells, showing median values and outliers up to 5.2 MW}
\label{fig:daily_distribution}
\end{figure}

The investigation demonstrates significant spatial and temporal variations in expected AC loads under the 35\% scenario. The hourly distribution analysis (Figure~\ref{fig:daily_distribution}) shows that while the median additional load across grid cells remains relatively low, substantial outliers emerge with peak values reaching 5.2 MW in individual 1km\textsuperscript{2} cells. These extreme loads are concentrated in densely populated metropolitan areas with high temperatures, particularly around large cities such as Karlsruhe, Braunschweig, and Frankfurt. A critical finding is the temporal concentration of these peak loads during evening hours, creating a concerning overlap with periods of typically higher baseline electricity demand and reduced solar PV electricity generation.

\begin{figure}[htbp]
\centering
\adjustbox{max width=\columnwidth}{
\includegraphics{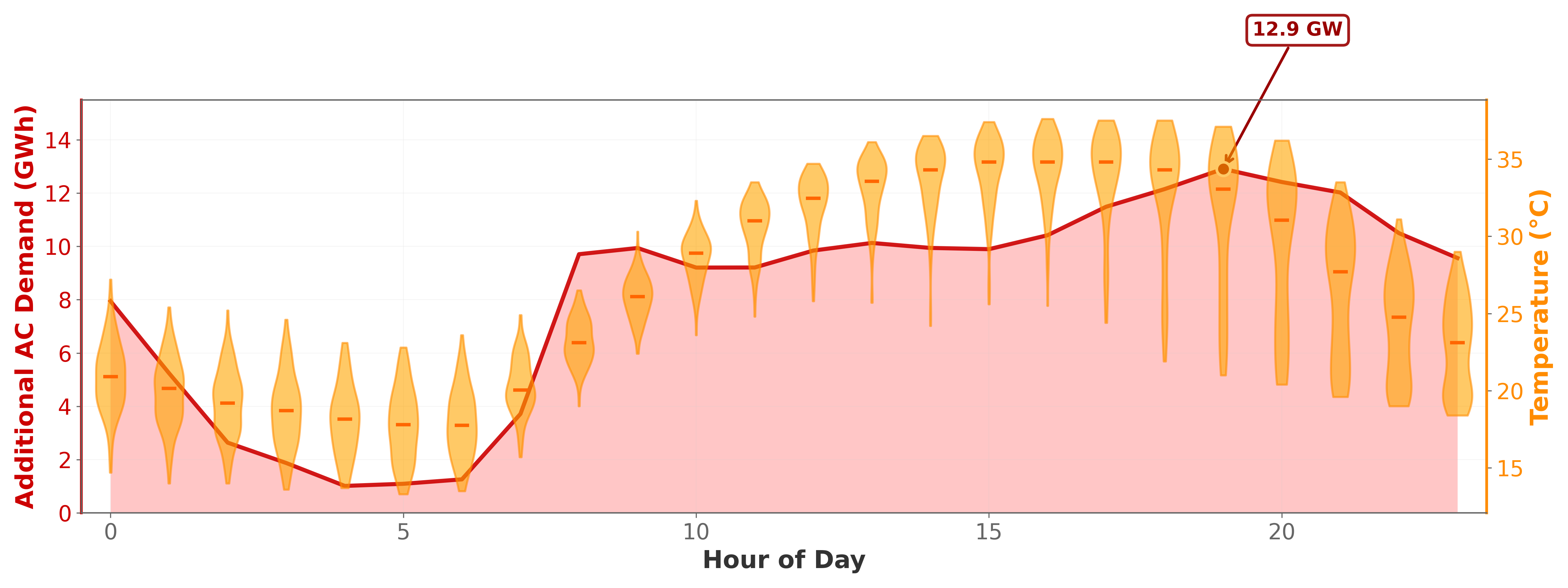}
}
\caption{Expected additional German electricity demand from mobile AC installations, peaking at 7 PM with 12.91 GWh}
\label{fig:germany_load}
\end{figure}

At the national level, our results indicate substantial aggregate demand impacts (Figure~\ref{fig:germany_load}). The expected additional German electricity load from mobile AC installations peaks at 7 PM with over 12.9 GW on the observed day. This represents a significant burden on the electricity system, particularly during evening peak hours when renewable generation capacity is typically lower. 

\begin{figure}[htbp]
\centering
\adjustbox{max width=\columnwidth}{
\includegraphics[width=0.8\columnwidth]{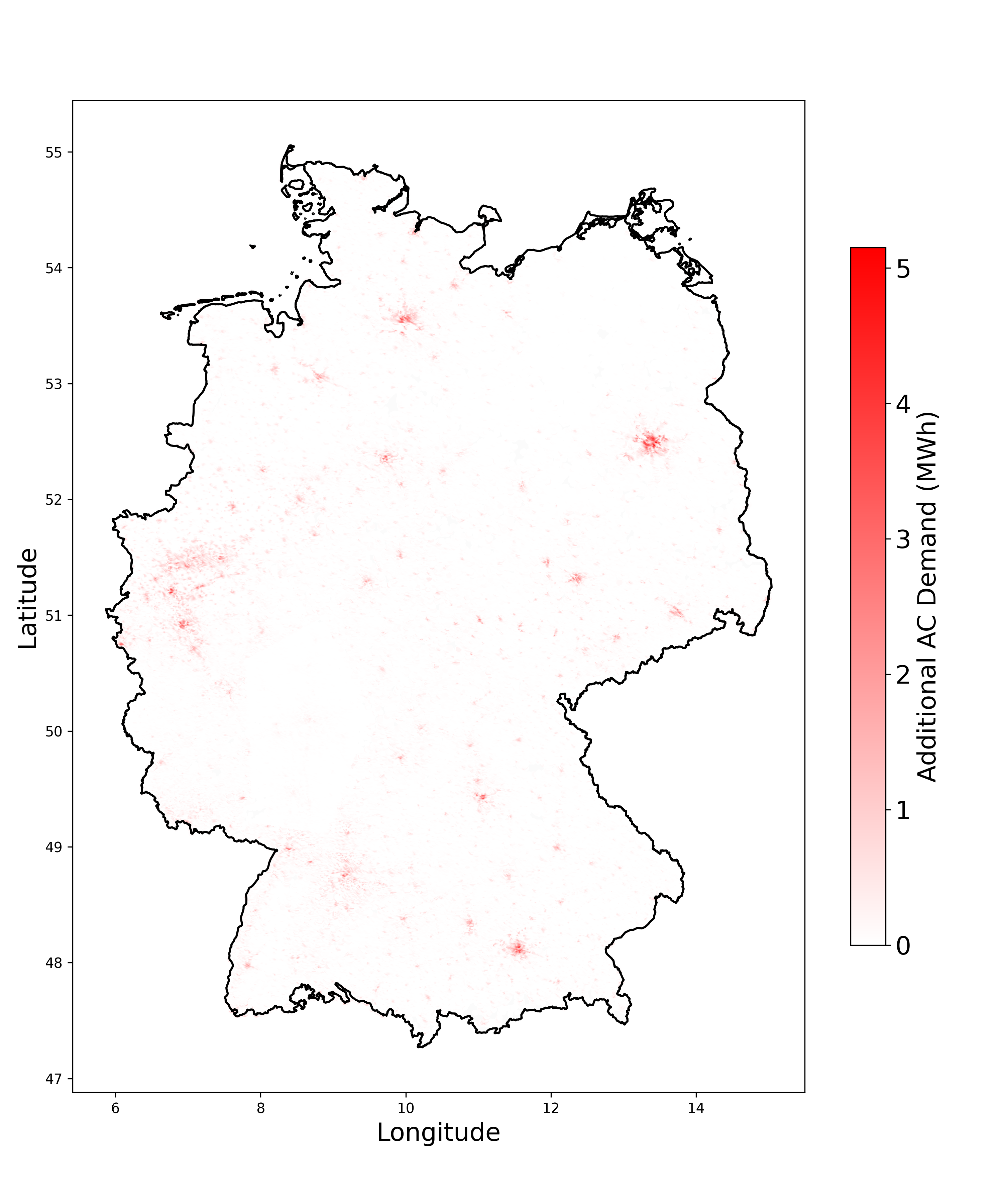}
}
\caption{Spatial distribution of additional AC load at peak hour (7 PM) across German grid cells, showing concentration in heavily populated areas}
\label{fig:ac_load_hour14}
\end{figure}

The spatial distribution analysis at peak load (Figure~\ref{fig:ac_load_hour14}) confirms that the highest additional demands are predominantly located in heavily populated urban areas, while rural and less populated regions show moderate effects. This heterogeneous distribution has important implications for grid planning, as transmission and distribution infrastructure in metropolitan areas may require substantial upgrades to accommodate these additional loads.

\begin{figure}[htbp]
\centering
\adjustbox{max width=\columnwidth}{
\includegraphics{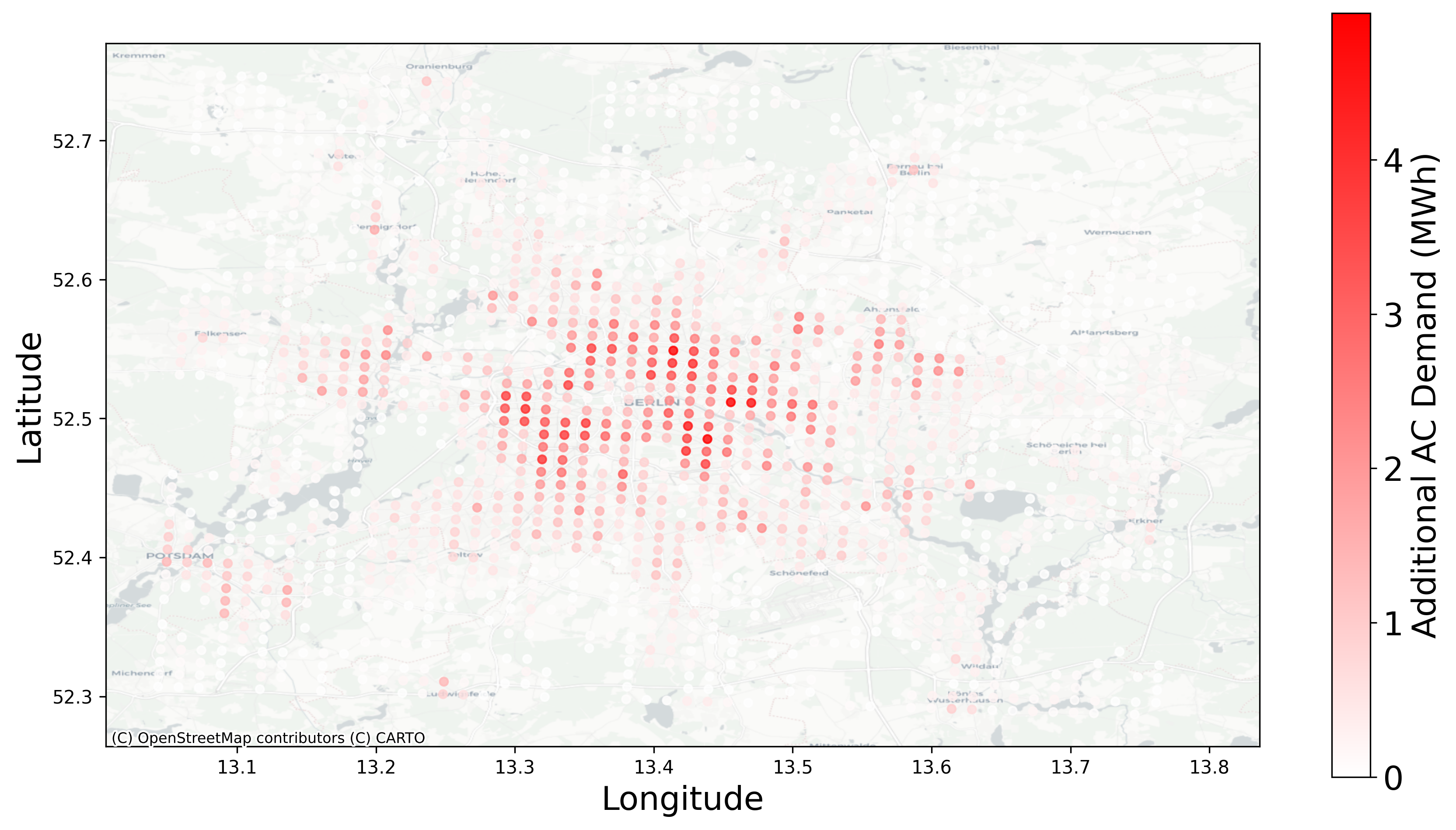}
    }
    \caption{Detailed view of AC load distribution in the area of Berlin, illustrating urban concentration effects}
    \label{fig:berlin_city_center}
\end{figure}

The detailed analysis of Berlin (Figure~\ref{fig:berlin_city_center}) exemplifies the urban concentration effect, showing how AC demand clusters in city centers where population density is highest. This localized analysis demonstrates the importance of city-level planning for AC-related electricity demand increases.

These findings highlight the need for proactive energy system planning to accommodate the growing demand for cooling systems driven by climate change. Especially, the afternoon demand peaks pose particular challenges for power system stability. Subsequently, we discuss the effects of these findings, potential remedies, and methodological limitations.

We note that the resulting load curves exhibit strong dependence on the underlying temperature-dependent activation functions. Our analysis employs parameters from \cite{liu2024exploring}, which assume AC activation at relatively low temperatures, as illustrated in Figure~\ref{fig:heterstoactiv} in the Appendix. To assess sensitivity, we conduct additional analysis using an activation function with relatively high temperature thresholds. Those results, presented in Figure~\ref{fig:aggregated_comp} in the Appendix, demonstrate a reduced peak demand of 9.5~GW. Notably, the timing of the AC demand peak remains consistent across both scenarios, confirming the directional validity of our findings. Future research should empirically investigate temperature-dependent air conditioning activation curves specific to German households to enable more precise modeling, as such country-specific empirical data remains unavailable.

\section{Discussion} \label{sec_discussion}

Our study gives a first conservative estimate of what would happen if approximately 20\% of households that have not yet adopted AC systems were to install mobile AC systems, increasing the total adoption rate from the current 19\% to 35\%: nationwide German electricity demand would increase by 12.91 GWh at 7 PM on the investigated 2nd of July 2025. Total electricity demand in Germany at this time was 61.3 GW \citep{smard}, so the AC-induced additional demand would represent a 21.1\% increase. Given the substantial potential increase, this would have consequences for utilities, grid operators, and trading firms, particularly in response to the novel demand peaks. We emphasize that our estimate is rather conservative, as it only considers a relatively small and efficient 2.1 kW AC device per household, a modest adoption rate of maximum one device per household, and neglects thermal inertia (see Subsection \ref{subsec_limitations}). Our proposed methodology enables involved stakeholders to develop well-reasoned estimates of AC-induced effects on electricity demand and provides the potential to investigate alternative AC equipment specifications, activation functions, and adoption rates.

A key finding of our study is an indication of \textit{when} additional AC loads can be expected: during afternoon and evening hours. This is caused by considering mobility patterns within demographic groups and real-world outdoor temperature curves. In practice, this is likely to be exacerbated even further by the thermal inertia of buildings during heatwaves \citep{zinzi2020thermal}. While we consider only outdoor temperatures as input for the activation function of the AC systems, households base their decisions on indoor temperature, which might be higher in evening hours when the home has been pre-heated and thermal energy is stored in the building mass \citep{liu2024exploring,zinzi2020thermal}. Therefore, evening electricity demand from AC systems will likely be higher than calculated in this study, which has implications for the operation of the power system, since this additional demand does not coincide with PV production in distribution systems. Locational incentives like dynamic grid charges and a further build-out of storage systems could be remedies for this.

\subsection{Limitations} \label{subsec_limitations}

Our analysis is subject to several important assumptions and simplifications that are important to take into consideration.

First, we focus exclusively on the residential sector, neglecting potential increases in commercial and industrial sectors where additional cooling demand can also be expected during heatwaves \citep{henze2007impact}. This limitation suggests our estimates may represent a conservative lower bound of total AC-related demand increases.

Second, within the residential sector, we consider only one type of AC system, i.e., a mobile AC system with relatively low energy consumption and high efficiency. In reality, many households may opt for more powerful HVAC system installations \citep{goetzler2016future}, especially in single-family homes, which would result in substantially higher electricity demands than our estimates suggest. Our assumption of efficient mobile units, therefore, likely underestimates the actual load increases.

Third, our methodology relies on simplified assumptions about mobility patterns and demographic distributions, applying uniform socio-demographic profiles across all grid cells with similar household size compositions. While these assumptions are based on reasonable demographic expectations, they may not capture the full heterogeneity of actual population distributions and behavioral patterns across different regions of Germany.

Fourth, we assume at most one AC system per household, regardless of household size or dwelling characteristics. However, larger households or multi-story residential buildings may install multiple AC units, potentially leading to significantly higher per-household electricity consumption than our estimates suggest.

Fifth, our approach neglects thermal building models and assumes that outdoor temperature directly corresponds to indoor temperature and AC activation needs. In this regard, we follow previous research, arguing that opening windows e.g., during the night at lower outdoor temperatures, has a resetting effect and therefore, outdoor temperature is an adequate heuristic for AC activation. Nevertheless, buildings with good insulation and fewer windows may maintain substantially lower indoor temperatures, especially during single hot days, reducing both AC installation likelihood and activation frequency. However, we already only model 35\% total AC adoption, which might already cover non-adoption by households with better thermal dynamics. Urban heat island effects, i.e., cities being unable to cool down at night compared to rural areas, will further exacerbate AC usage in cities and thus lead to even more pronounced concentration effects than we already find \citep{parker2010urban}.

Finally, we model AC operation as binary (either fully on or completely off) rather than considering variable-speed units that can operate at intermediate power levels \citep{ma2009energy}. Modern AC systems often include variable-speed compressors and fans that can adjust output based on cooling demand, potentially operating at 50\% or other intermediate power levels. While our model's temperature-dependent activation function somewhat accounts for different temperatures, a non-binary model may be more precise, especially during non-peak hours.

\subsection{Policy Implications} \label{subsec_policy}

Our findings reveal substantial implications for energy policy and grid planning that require immediate attention from policymakers and system operators. The rapid adoption potential of mobile AC systems presents unique challenges that differ significantly from traditional energy infrastructure planning timelines.

The most critical concern is the speed at which mobile AC adoption could occur. Unlike conventional HVAC systems that require professional installation, mobile AC units can be purchased and deployed within hours of a heatwave onset. During multi-day extreme heat events, thousands of households may simultaneously decide to install these systems, creating sharp and unexpected load increases that could overwhelm unprepared grid infrastructure. Moreover, whereas larger HVAC systems and electric vehicles are sometimes directly controllable by grid operators, mobile AC systems typically are not. For example, in Germany, only HVAC systems above 4.2 kW are defined as controllable loads \citep{bnetza_2023}. This necessitates proactive rather than reactive policy measures.

Policymakers must address two critical dimensions of this challenge. First, peak demand management requires careful consideration of temporal load patterns. The midday component of AC demand coincides with photovoltaic production peaks, suggesting that continued PV build-out can partially offset daytime cooling loads, also on a local level. However, the pronounced afternoon peak of over 12.9 GW occurs during periods of reduced solar generation. Counteractions from policymakers could include incentivizing technological flexibility options (like battery storage systems) or demand-response-based options (like dynamic tariffs).

Second, urban grid infrastructure requires strategic future-proofing investments. Metropolitan areas face the highest risk of localized grid overloading, with individual grid cells potentially experiencing load increases of up to 5.2 MW. Distribution networks in densely populated areas may require substantial upgrades to accommodate these concentrated demand increases, particularly in aging urban infrastructure that was not designed for such rapid load growth.

For areas with a high risk of urban grid overloading, policymakers should consider several mitigation pathways. Balcony photovoltaic installations that are especially suitable for multi-family homes in urban areas could provide targeted peak load reduction while simultaneously reducing electricity costs for lower-income households who may be disproportionately affected by AC operation expenses. Additionally, policies promoting more efficient AC systems through subsidies and rebates could substantially reduce peak demand effects while maintaining cooling comfort.

Finally, longer-term urban planning strategies should incorporate building insulation improvements and broader city cooling concepts such as increased urban greenery, reflective building materials, and strategic urban design to reduce overall cooling demand. These comprehensive approaches address the root causes of excessive urban heat rather than merely responding to its symptoms through electric AC.

\section{Conclusion} \label{sec_conclusion}

This study provides a novel method for the spatially and temporally disaggregated analysis of electricity demand increases from mobile AC systems, as well as the first corresponding case study for Germany. It reveals significant implications for energy system planning and policy development. Our investigation of spatially disaggregated grid cells demonstrates that increasing mobile AC adoption from the current 19\% to 35\% - as expected for 2030 - could result in peak additional demand of over 12.9 GW, with substantial spatial concentration in urban metropolitan areas.

The analysis reveals two critical findings that demand immediate policy attention. First, the temporal pattern of AC demand creates a pronounced afternoon peak that coincides with reduced photovoltaic generation, potentially exacerbating power system stability challenges during already critical peak demand periods. Second, the spatial concentration of extreme loads in densely populated areas, with individual 1km\textsuperscript{2} grid cells experiencing up to 5.2 MW increases, poses significant risks for urban distribution infrastructure. In summary, the rapid deployment potential of mobile AC systems creates unprecedented challenges for grid planning, as thousands of units could be installed within hours of a heatwave onset, generating sharp and unexpected demand spikes.

Our findings underscore the urgent need for proactive energy system adaptation strategies that address both immediate power system stability concerns and long-term climate resilience. The convergence of climate change impacts, rapid technology adoption, and aging infrastructure may create a perfect storm that requires coordinated responses across multiple policy domains. Distribution grid operators must prepare for sudden load increases during extreme weather events, while policymakers must develop comprehensive strategies that may include targeted photovoltaic deployment, battery storage expansion, and urban infrastructure upgrades.

While our analysis focuses on Germany, the methodology and findings have broader relevance for other geographies facing similar climate challenges and energy system transitions. The rapid adoption potential of mobile cooling technologies represents a new category of demand that might be less controllable for distribution grid operators than heat pumps and electric vehicles.

Future research should address the limitations identified in this study, particularly the integration of thermal building models, commercial sector demand, and detailed mobility patterns. Additionally, the development of real-time monitoring systems and demand response programs designed explicitly for cooling loads will be essential for managing the transition to higher AC adoption rates while maintaining power system stability and affordability.

The challenge of climate-driven electricity demand increases is not merely a technical problem but a comprehensive policy challenge. The window for proactive preparation is rapidly closing, making immediate action essential for ensuring power system stability in the face of increasing extreme heat events.

\section*{Declaration of generative AI in scientific writing}
During the preparation of this work, the authors used Grammarly, ChatGPT 4o and 5, and Claude Sonnet 4 for language refinement and grammar correction. 
After using these tools, the authors reviewed and edited the content as needed and take full responsibility for the publication's content.

\section*{Declaration of competing interest}

The authors declare that they have no known competing financial interests or personal relationships that could have appeared to influence the work reported in this paper.

\bibliography{sample}

\begin{thebibliography}{}

\bibitem[{B. Clarke et al.}, 2025]{heat-deaths-2025}
{B. Clarke et al.} (2025).
\newblock {Climate change tripled heat-related deaths in early summer European heatwave. Grantham Institute report}.

\bibitem[Bundesnetzagentur, 2025]{smard}
Bundesnetzagentur (2025).
\newblock {SMARD - Strommarktdaten}.
\newblock \url{https://www.smard.de/home}.
\newblock Accessed: 2025-07-10.

\bibitem[{Bundesnetzagentur, Beschlusskammer 6}, 2023]{bnetza_2023}
{Bundesnetzagentur, Beschlusskammer 6} (2023).
\newblock Beschluss vom 27.11.2023.
\newblock \url{https://www.bundesnetzagentur.de/DE/Beschlusskammern/1_GZ/BK6-GZ/2022/BK6-22-300/Beschluss/BK6-22-300_Beschluss_20231127.pdf?}

\bibitem[Colelli et~al., 2023]{colelli2023air}
Colelli, F.~P., Wing, I.~S., and Cian, E.~D. (2023).
\newblock Air-conditioning adoption and electricity demand highlight climate change mitigation--adaptation tradeoffs.
\newblock {\em Scientific reports}, 13(1):4413.

\bibitem[Dahlström et~al., 2022]{DAHLSTROM2022112099}
Dahlström, L., Broström, T., and Widén, J. (2022).
\newblock Advancing urban building energy modelling through new model components and applications: A review.
\newblock {\em Energy and Buildings}, 266:112099.

\bibitem[{De'Longhi Appliances S.r.l.}, 2025]{delonghi_pac_es72_young}
{De'Longhi Appliances S.r.l.} (2025).
\newblock Pinguino compact tragbare klimaanlage pac es72 young.
\newblock \url{https://www.delonghi.com/de-de/p/mobile-klimagerate-pinguino-compact-tragbare-klimaanlage-pac-es72-young/PACES72YOUNG.html?pid=0151453006}.
\newblock Accessed: 2025-07-10.

\bibitem[{Federal Statistical Office (Destatis)}, 2024]{zensus2022_population}
{Federal Statistical Office (Destatis)} (2024).
\newblock 2022 census: Germany’s population at 82.7 million.
\newblock \url{https://www.zensus2022.de/EN/News/2022_census_Germanys_population_at_82.7_million.html}.
\newblock Press Release No. 44, June 26, 2024.

\bibitem[Ferrando et~al., 2020]{Ferrando_2020}
Ferrando, M., Causone, F., Hong, T., and Chen, Y. (2020).
\newblock Urban building energy modeling (ubem) tools: A state-of-the-art review of bottom-up physics-based approaches.
\newblock {\em Sustainable Cities and Society}, 62:102408.

\bibitem[Fischer and Sch{\"a}r, 2010]{fischer2010consistent}
Fischer, E.~M. and Sch{\"a}r, C. (2010).
\newblock Consistent geographical patterns of changes in high-impact european heatwaves.
\newblock {\em Nature geoscience}, 3(6):398--403.

\bibitem[Flett and Kelly, 2021]{flett2021modelling}
Flett, G. and Kelly, N. (2021).
\newblock Modelling of individual domestic occupancy and energy demand behaviours using existing datasets and probabilistic modelling methods.
\newblock {\em Energy and Buildings}, 252:111373.

\bibitem[Fukami et~al., 2022]{Japan_2022}
Fukami, R., Hagishima, A., Tanimoto, J., and Ikegaya, N. (2022).
\newblock Stochastic nature of occupants' behavior toward air-conditioning operation in residential buildings.
\newblock {\em Japan Architectural Review}, 5(4):649--660.

\bibitem[{Galaxus}, 2025]{Galaxus_2025}
{Galaxus} (2025).
\newblock {Klimaanlagen: Hitzewelle sorgt für Rekordverkäufe}.
\newblock \url{https://www.galaxus.de/de/page/klimaanlagen-hitzewelle-sorgt-fuer-rekordverkaeufe-38590}.
\newblock Accessed: 2025-08-15.

\bibitem[Goetzler et~al., 2016]{goetzler2016future}
Goetzler, W., Guernsey, M., Young, J., Fujrman, J., and Abdelaziz, A. (2016).
\newblock The future of air conditioning for buildings.
\newblock Technical report, Navigant Consulting, Burlington, MA (USA).

\bibitem[Gunay et~al., 2018]{Gunay02012018}
Gunay, H.~B., O'Brien, W., Beausoleil-Morrison, I., and Bursill, J. (2018).
\newblock Development and implementation of a thermostat learning algorithm.
\newblock {\em Science and Technology for the Built Environment}, 24(1):43--56.

\bibitem[Henze et~al., 2007]{henze2007impact}
Henze, G.~P., Pfafferott, J., Herkel, S., and Felsmann, C. (2007).
\newblock Impact of adaptive comfort criteria and heat waves on optimal building thermal mass control.
\newblock {\em Energy and Buildings}, 39:21--235.

\bibitem[Huber et~al., 2019]{huber_2019}
Huber, J., Höffer, J., Thumm, J., and Weinhardt, C. (2019).
\newblock {Parking events derived from trip data from MiD2008}.
\newblock Karlsruhe Institute of Technology, \url{https://doi.org/10.35097/1191}.

\bibitem[{IEA}, 2025]{IEA}
{IEA} (2025).
\newblock Vglobal energy review 2025.
\newblock \url{https://www.iea.org/reports/global-energy-review-2025}.

\bibitem[Liu et~al., 2024a]{liu2024rising}
Liu, J., Qi, J., Yin, P., Liu, W., He, C., Gao, Y., Zhou, L., Zhu, Y., Kan, H., Chen, R., and Zhou, M. (2024a).
\newblock Rising cause-specific mortality risk and burden of compound heatwaves amid climate change.
\newblock {\em Nature Climate Change}, 14(11):1201--1209.

\bibitem[Liu et~al., 2024b]{liu2024exploring}
Liu, Z., Dou, Z., Chen, H., Zhang, C., Wang, S., Wu, Y., Liu, X., and Yan, D. (2024b).
\newblock Exploring the impacts of heterogeneity and stochasticity in air-conditioning behavior on urban building energy models.
\newblock {\em Sustainable Cities and Society}, 103:105285.

\bibitem[Ma and Wang, 2009]{ma2009energy}
Ma, Z. and Wang, S. (2009).
\newblock Energy efficient control of variable speed pumps in complex building central air-conditioning systems.
\newblock {\em Energy and Buildings}, 41(2):197--205.

\bibitem[Neubauer, 2024]{verivox2024klimaanlagen}
Neubauer, L. (2024).
\newblock {Immer mehr Deutsche nutzen Klimaanlagen}.
\newblock \url{https://www.verivox.de/strom/nachrichten/immer-mehr-deutsche-nutzen-klimaanlagen-1120788/}.
\newblock Accessed: 2025-07-13.

\bibitem[{Nürnberger Nachrichten}, 2025]{nn2025klimageraete}
{Nürnberger Nachrichten} (2025).
\newblock {Klimaanlagen und Ventilatoren in Roth und Schwabach ausverkauft}.
\newblock \url{https://www.nn.de/region/roth/klimaanlagen-und-ventilatoren-in-roth-und-schwabach-ausverkauft-hier-gibt-es-noch-gerate-1.14744590}.
\newblock Accessed: 2025-07-10.

\bibitem[Parker, 2010]{parker2010urban}
Parker, D.~E. (2010).
\newblock Urban heat island effects on estimates of observed climate change.
\newblock {\em Wiley Interdisciplinary Reviews: Climate Change}, 1(1):123--133.

\bibitem[Ren et~al., 2014]{ren_2014}
Ren, X., Yan, D., and Wang, C. (2014).
\newblock Air-conditioning usage conditional probability model for residential buildings.
\newblock {\em Building and Environment}, 81:172--182.

\bibitem[Staffell et~al., 2023]{ninja_2023}
Staffell, I., Pfenninger, S., and Johnson, N. (2023).
\newblock A global model of hourly space heating and cooling demand at multiple spatial scales.
\newblock {\em Nature Energy}, pages 1328--1344.

\bibitem[{Statistisches Bundesamt}, 2022a]{zensus2022_table1035}
{Statistisches Bundesamt} (2022a).
\newblock {Ergebnisse Zensus 2022 -- Table 1000A-1035}.
\newblock \url{https://ergebnisse.zensus2022.de/datenbank/online/statistic/1000A/table/1000A-1035}.

\bibitem[{Statistisches Bundesamt}, 2022b]{zensus2022_table2081}
{Statistisches Bundesamt} (2022b).
\newblock {Ergebnisse Zensus 2022 -- Table 1000A-2081}.
\newblock \url{https://ergebnisse.zensus2022.de/datenbank/online/statistic/1000A/table/1000A-2081}.

\bibitem[{SWR Aktuell}, 2025]{swr_2025}
{SWR Aktuell} (2025).
\newblock {Pfalz: Ventilatoren und mobile Klima-Anlagen heiß begehrt}.
\newblock \url{https://www.swr.de/swraktuell/rheinland-pfalz/ludwigshafen/ventilatoren-und-mobile-klimageraete-in-der-pfalz-ausverkauft-100.html}.
\newblock Accessed: 2025-07-10.

\bibitem[{Visual Crossing Corporation}, 2025]{visualcrossing_weather}
{Visual Crossing Corporation} (2025).
\newblock Visual crossing weather (2025).
\newblock \url{https://www.visualcrossing.com/}.

\bibitem[Wierda and Zanuttini, 2024]{eu_commission_ac_estimates_2030}
Wierda, L. and Zanuttini, A. (2024).
\newblock Ecodesign impact accounting. overview report 2024.
\newblock Technical report, European Commission.

\bibitem[Zhang et~al., 2024]{zhang2024climate}
Zhang, M., Ma, X., Wang, W., Sheng, J., Cao, J., Cheng, Z., and Zhang, X. (2024).
\newblock Climate adaptation investments: Short-term shocks and long-term effects of temperature variation on air conditioning adoption.
\newblock {\em Sustainable Cities and Society}, 108:105493.

\bibitem[Zinzi et~al., 2020]{zinzi2020thermal}
Zinzi, M., Agnoli, S., Burattini, C., and Mattoni, B. (2020).
\newblock On the thermal response of buildings under the synergic effect of heat waves and urban heat island.
\newblock {\em Solar Energy}, 211:1270--1282.

\end{thebibliography}

\clearpage
\section*{Appendix}
\label{ref:appendix}

\subsection*{Weather stations}
\begin{figure}[htbp]
\centering
\adjustbox{max width=\columnwidth}{
\includegraphics{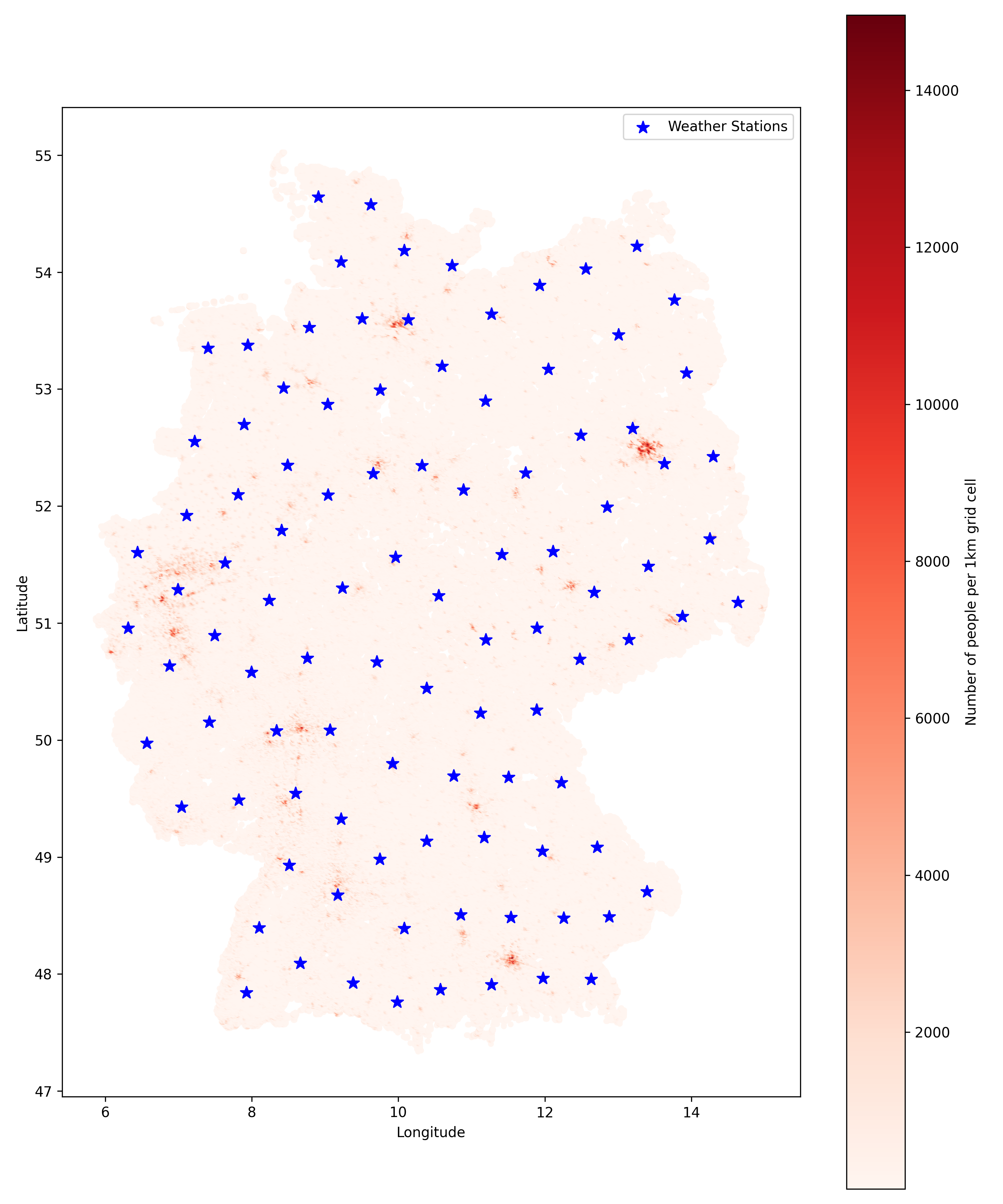}
}
\caption{Geographic distribution of weather stations}
\label{fig:weather_stations}
\end{figure}

In Figure \ref{fig:weather_stations}, the 100 weather stations from which temperature measurements were retrieved are depicted. Grid cells were matched to the closest weather stations based on a k-means clustering approach.

\subsection*{Household Composition Calculations}
The household compositions are calculated based on German census data from 2022, based on tables of household sizes and inhabitants \citep{zensus2022_table2081} and pensioners statistics \citep{zensus2022_table1035}. Tab. \ref{tab:household_composition} displays the used household data. The detailed calculations can be found in the following:

\begin{itemize}
    \item \textbf{1-Person Households:} Out of 17,075,514 single households, approximately 35\% (6,026,872) are retired seniors, thus approximately 35\% retired and 65\% other singles.
    
   \item \textbf{2-Person Households:} Of 24,390,153 two-person households, approximately 31\% are retired couples without children (7,466,286), 47\% are non-retired couples without children (11,530,424), 15\% are single parents (3,641,070 families), and 7\% are shared flats (1,752,375).

    \item \textbf{3-Person Households:} Out of 14,798,064 three-person households, approximately 74\% are couples with children (10,909,206 families), 15\% are single-parent families (2,282,370 families), 8\% are couples without children (1,226,304), and 3\% are shared flats (380,193).
    
    \item \textbf{4-Person Households:} For 14,560,002 four-person households, approximately 89\% are couples with children (12,944,656 families), 7\% are single-parent families (999,176 families), 3\% are couples without children (454,858), and 1\% are shared flats (161,316).
    
    \item \textbf{5-Person Households:} Out of 5,947,995 five-person households, approximately 90\% are couples with children (5,350,339 families), 6\% are single-parent families (380,111), 2\% are couples without children (145,675), and 1\% are shared flats (71,867).
    
    \item \textbf{6+ Person Households:} For 4,450,086 households with six or more persons, approximately 84\% are couples with children (3,732,121 families), 6\% are single-parent families (284,842), 6\% are couples without children (268,374), and 4\% are shared flats (164,751).
\end{itemize}

We note that the calculations include some simplifications. For instance, there are 131,314 pensioners living in households in sizes of three or more. Since it is a small portion of pensioners (0.2\%) and cannot be allocated to exact household sizes, we neglect the number.

\begin{table*}[h]
\centering
\resizebox{\textwidth}{!}{
\begin{tabular}{lrrrrr}
\hline
\textbf{Household Type (Persons)} & \textbf{Families \%} & \textbf{Couples without children \%} & \textbf{Retired \%} & \textbf{Shared flats \%} & \textbf{Singles \%} \\
\hline
1-Person & 0 & 0 & 35 & 0 & 65 \\
2-Person & 15 & 47 & 31 & 7 & 0 \\
3-Person & 89 & 8 & 0 & 3 & 0 \\
4-Person & 96 & 3 & 0 & 1 & 0 \\
5-Person & 96 & 2 & 0 & 1 & 0 \\
6+ Person & 90 & 6 & 0 & 4 & 0 \\
\hline
\end{tabular}
}
\caption{Household composition by household size in percent.}
\label{tab:household_composition}
\end{table*}

\subsection*{AC activation function}

\begin{figure}[htbp]
\centering
\adjustbox{max width=\columnwidth}{
\includegraphics{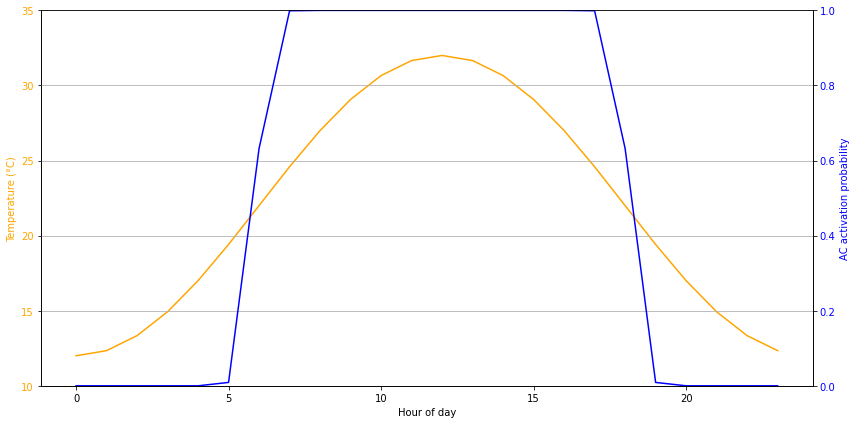}
}
\caption{AC activation probability at exemplary temperatures.}
\label{fig:heterstoactiv}
\end{figure}

In Figure \ref{fig:heterstoactiv}, we show how the AC activation probability (based on Weibull cumulative function from \cite{liu2024exploring}, parametrized with threshold temperature $u = 18.5^\circ$C, scale parameter $l = 3.5^\circ$C and shape parameter $k = 3.5$) behaves at different temperature levels.

\subsection*{Sensitivity analysis}

To assess the robustness of our findings, we conducted a sensitivity analysis using alternative parameters for the AC activation function. While our main analysis employed some of the Heter\_Sto parameters from \cite{liu2024exploring} with threshold temperature $u = 18.5°C$, scale parameter $l = 3.5°C$, and shape parameter $k = 3.5$, the sensitivity analysis uses Homo\_Sto parameters with higher activation thresholds: $u = 25.5°C$, $l = 6.5°C$, and $k = 6.5$.

\begin{figure}[htbp]
\centering
\adjustbox{max width=\columnwidth}{
\includegraphics{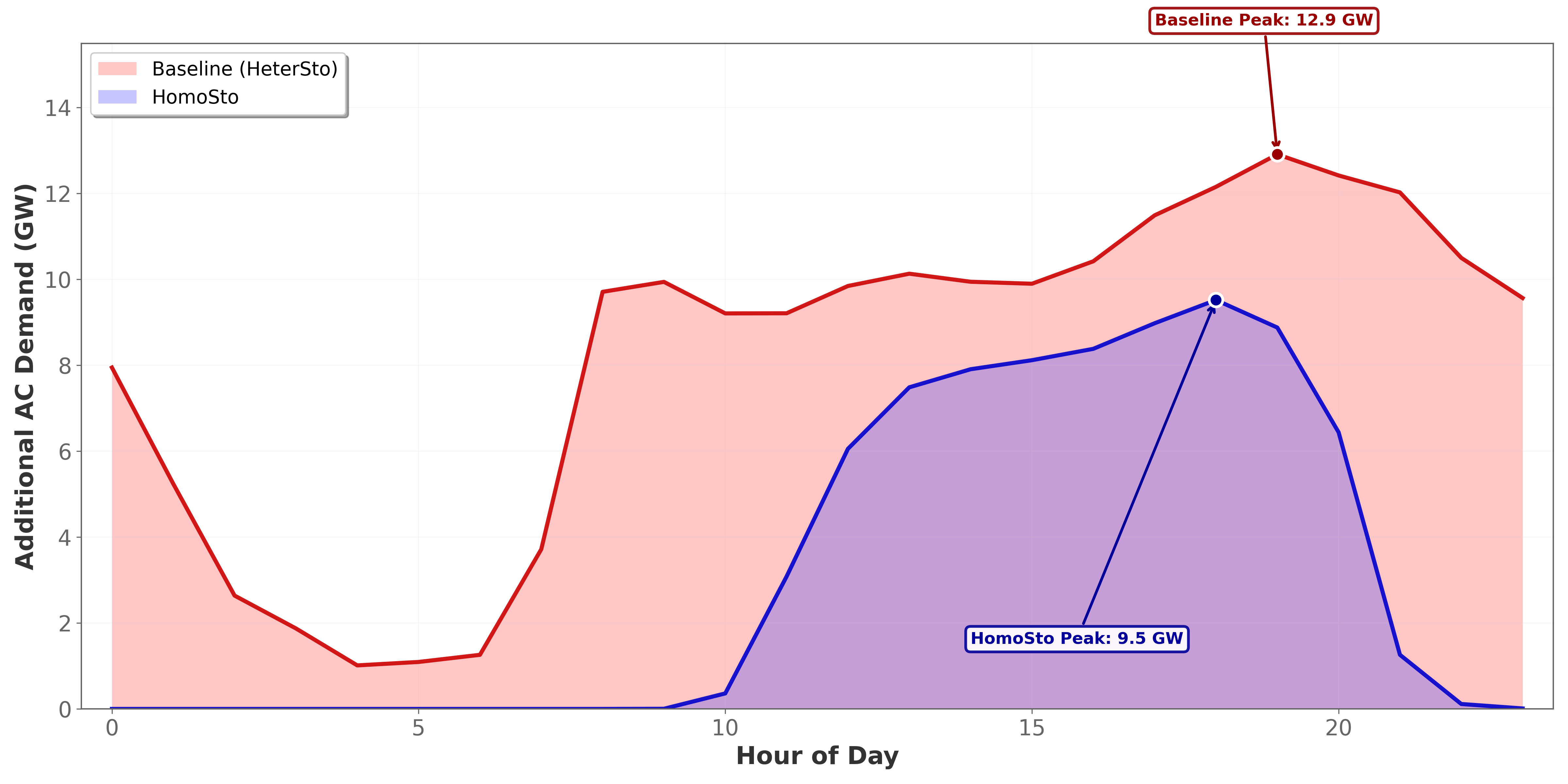}
}
\caption{Sensitivity analysis aggregated load}
\label{fig:aggregated_comp}
\end{figure}

Figure \ref{fig:aggregated_comp} compares the aggregated national electricity demand between both scenarios. The higher temperature threshold scenario (Homo\_Sto) results in a reduced peak demand of approximately 9.5 GW compared to 12.9 GW in the baseline case. This represents a 26\% reduction in peak load, demonstrating the significant influence of activation function parameters on demand estimates. Importantly, the temporal pattern remains consistent across both scenarios, with peak activation occurring during afternoon and evening hours, confirming the directional validity of our timing-related findings.

\begin{figure}[htbp]
\centering
\adjustbox{max width=\columnwidth}{
\includegraphics{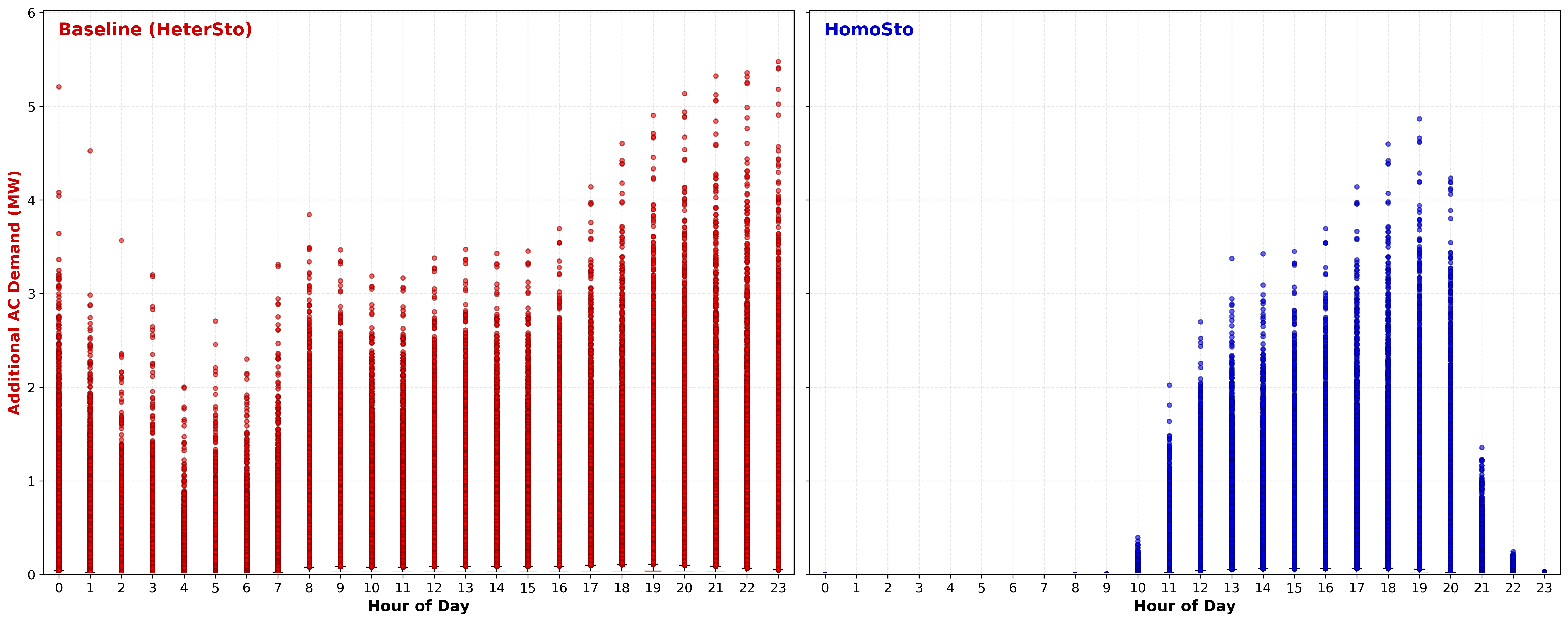}
}
\caption{Sensitivity analysis distribution}
\label{fig:distribution_comp}
\end{figure}

The spatial distribution analysis (Figure \ref{fig:distribution_comp}) reveals that while absolute demand levels differ substantially between scenarios, the relative concentration patterns remain similar. The higher threshold scenario shows maximum loads per grid cell reaching 4.2 MW compared to 5.2 MW in the baseline case.

\begin{figure}[htbp]
\centering
\adjustbox{max width=\columnwidth}{
\includegraphics{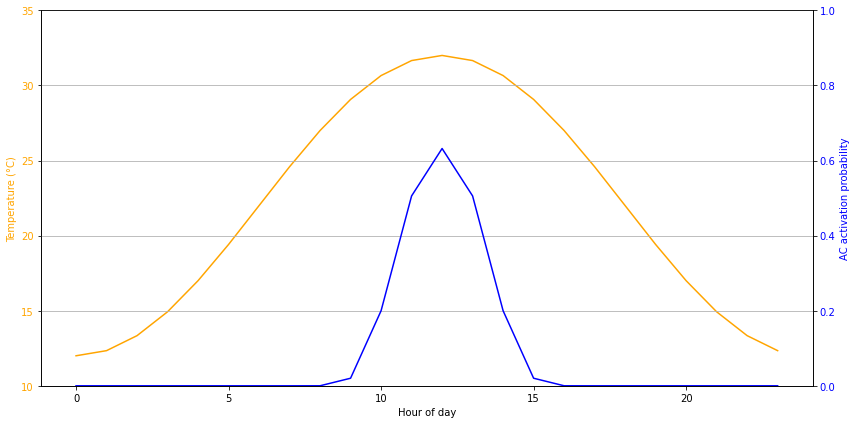}
}
\caption{Homo\_Sto AC activation probability at exemplary temperatures, used for sensitivity analysis.}
\label{fig:homostoactiv}
\end{figure}

\end{document}